\documentclass[aps,prx,twocolumn,superscriptaddress,nofootinbib,notitlepage]{revtex4-1}
\usepackage{graphicx}
\usepackage{amssymb}
\usepackage{hyperref}
\usepackage{amsmath}
\usepackage{amsfonts}
\usepackage{subfigure}
\usepackage{dsfont}
\usepackage{dcolumn}
\usepackage{esint}
\usepackage{mathrsfs}
\usepackage{bm}
\usepackage{multirow}
\usepackage{color}
\usepackage{datetime}

\begin{document}
\title{Quantum Entanglement of Non-Hermitian Quasicrystals}

\author{Li-Mei Chen}
\thanks{These authors contributed equally to this work.}
\affiliation{School of Physics, Sun Yat-sen University, Guangzhou, 510275,
China}
\author{Yao Zhou}
\thanks{These authors contributed equally to this work.}
\affiliation{School of Physics, Sun Yat-sen University, Guangzhou, 510275,
China}
\author{Shuai A. Chen}
\email{chsh@ust.hk}
\affiliation{Department of Physics, The Hong Kong University of Science and Technology, Hong Kong SAR, China}
\affiliation{Institute for Advanced Study, Tsinghua University, Beijing,
100084, China}
\author{Peng Ye}
\email{yepeng5@mail.sysu.edu.cn}
\affiliation{School of Physics, Sun Yat-sen University, Guangzhou, 510275,
China}
\affiliation{State Key Laboratory of Optoelectronic Materials and
Technologies, Sun Yat-sen University, Guangzhou, 510275, China}

\date{\today}

\begin{abstract}
  As a hallmark of pure quantum effect, quantum entanglement  has  provided unconventional routes  to condensed matter systems. Here, from the perspective of quantum entanglement, we disclose exotic quantum physics in non-Hermitian quasicrystals. We study a class of experimentally realizable models for non-Hermitian quasicrystal chains, in which   asymmetric hopping and complex potential coexist. We diagnose global phase diagram by means of entanglement from both real-space and momentum-space partition. By measuring entanglement entropy, we numerically determine the metal-insulator transition point. We combine real-space and momentum-space entanglement spectra to complementarily characterize the delocalization phase and the localization phase. Inspired by entanglement spectrum, we further analytically prove that a duality exists between the two phase regions. The transition point is self-dual and   exact,  further validating the numerical result from diagonalizing non-Hermitian matrices. Finally, we identify mobility edge by means of entanglement.
  \end{abstract}
\maketitle

{\color{blue}\emph{Introduction.}}---Recently, non-Hermitian systems~\cite{ashidaNonHermitian2020,bergholtzExceptional2021}   have attracted increasing interests. Many striking properties, especially from the perspective of topological physics, have been explored theoretically and   experimentally, such as the generalized bulk-boundary correspondence~\cite{yaoEdge2018,yokomizoNonBloch2019}, the non-Hermitian skin effect~\cite{yaoEdge2018,Brandenbourger2019nonreciprocal,Zhang2021Observation} and the exceptional point~\cite{Hahn2016Observation,bergholtzExceptional2021}. While many studies focus on crystalline systems, non-Hermiticity has also been introduced to noncrystalline systems, e.g., quasicrystal systems~\cite{jagannathanFibonacci2021} and disorder systems~\cite{eversAnderson2008}.   As a paradigmatic quasicrystal lattice model, the celebrated Aubry-Andr\'e-Harper~(AAH) model~\cite{aubry1980analyticity, harperSingle1955} provides an example of   Anderson localization~\cite{eversAnderson2008} without disorder. Its delocalization-localization transition can be deduced from a self-duality argument. The absence of mobility edge is one of particular features of the AAH model. In other words, upon varying the quasi-periodic potential, all extended eigenstates simultaneously become exponentially localized. Moreover, in   non-Hermitian disorder systems, Hatano and Nelson~\cite{hatanoLocalization1996,hatanoVortex1997,hatanoNonHermitian1998} discovered that   localized states can be delocalized by the nonreciprocal hopping, which is also generalized to non-Hermitian quasicrystal systems  in recent years~\cite{jazaeriLocalization2001,gongTopological2018,longhiTopological2019,jiangInterplay2019,zengWinding2020,caiBoundarydependent2021,liuLocalization2021}.

To unveil exotic   non-Hermitian quantum effect, many theoretical approaches originally introduced in Hermitian quantum systems have been borrowed. Quantum entanglement is such an example~\cite{amicoEntanglement2008,ciracMatrix2021,changEntanglement2020, herviouEntanglement2019, chenEntanglement2020,leePositionmomentum2014,leeFreefermion2015,guoEntanglement2021,bacsiDynamics2021,okumaQuantum2021,sayyadEntanglement2021}. Practically, entanglement entropy (EE) \cite{eisertColloquium2010,islamMeasuring2015}   and entanglement spectrum (ES)~\cite{liEntanglement2008,turnerEntanglement2010}  can be  obtained   by partitioning Hilbert space into several subregions. The Hilbert space can be written in various representations, such as  real-space, momentum-space\cite{thomaleNonlocal2010,mondragonshemCharacterizing2013,mondragonshemSignatures2014}, and orbital-space. While EE is a single number, it is generically believed that ES contains more information of the underlying physical systems.       It has been  known that entanglement plays important roles in diagnosing exotic phases of matter in Hermitian quantum systems, e.g.,  topological orders \cite{Levin2006detecting,kitaevTopological2006} and   disorder systems~\cite{apollaroEntanglement2006,jiaEntanglement2008,Prodan2010Entanglement,berkovitsEntanglement2012,mondragonshemCharacterizing2013,mondragonshemSignatures2014,hamazakiNonHermitian2019}.       The aim of this work is   to introduce the entanglement approach and explore    non-Hermitian quasicrystals via EE and ES of both real-space and momentum-space partitions.

In this work, we propose a class of   1D quasicrystal models, in which two sources of non-Hermiticity (asymmetric hopping and complex energy potential) coexist. The feature of quasicrystal is induced by the presence of an irrational number $\alpha$ in complex energy potential. Then, by measuring EE, we numerically determine the transition point of metal-insulator transition (MIT). Noticing that, the non-Hermitian density matrix $\rho$ adopted here involves both left- and right-many-body states in bi-orthogonal basis.  Once the MIT point is determined, we combine the real-space and momentum-space ES to identify and explore the phase regions of both localization  and   delocalization. Inspired by the numerical results, we analytically prove  that, in terms of entanglement spectrum in both momentum-space and real-space, the delocalization phase and localization phase are      dual to each other. Meanwhile, the MIT   point is analytically exact and self-dual. This analytic result validates the numerical result of non-Hermitian matrix diagonalization. In the end, we discuss the physics of mobility edges from the perspective of entanglement.

{\color{blue}\emph{Model.}}---We start with a non-Hermitian 1D quasicrystal model   with   both the asymmetric hopping and the incommensurate complex potential:
\begin{equation}
\begin{split}
H=\sum_n\left(J_Rc_{n+1}^{\dag}c_{n}+J_Lc_n^{\dag}c_{n+1}\right)+\sum_n\mathcal{V}_nc_n^{\dag}c_n ~,
\end{split}
\label{model1}
\end{equation}
where $c^{\dagger}_{n} (c_{n})$ is the creation (annihilation) operator of spinless fermion at lattice site $n$. $\mathcal{V}_n=V\exp{(-2\pi i \alpha n)}$ is a site-dependent incommensurate complex potential with irrational number $\alpha$. The potential strength $V$ is a positive real number. The ingredients of this model~(\ref{model1}) are simple and can be realized in acoustic experiments~\cite{Brandenbourger2019nonreciprocal,chen2021efficient,Zhang2021Observation}. In the actual numerical simulations, we   approximately set the irrational number $\alpha$ as a rational number $\frac{M_{k}}{N_{k}}$ for $M_{k}$ and $N_{k}$ coprime, which has been generally adopted in the literature of quasicrystals (see  Supplemental Material~\cite{supmat}). Meanwhile, to satisfy the periodic boundary condition (PBC), we should adopt the system size $L=N_{k}$. Without loss of generality, we set $J_R=1 > J_L$ and $\alpha=\sqrt{2}\approx \frac{239}{169}$.

\begin{figure}[htbp]
\centering
\includegraphics[width=8.5cm]{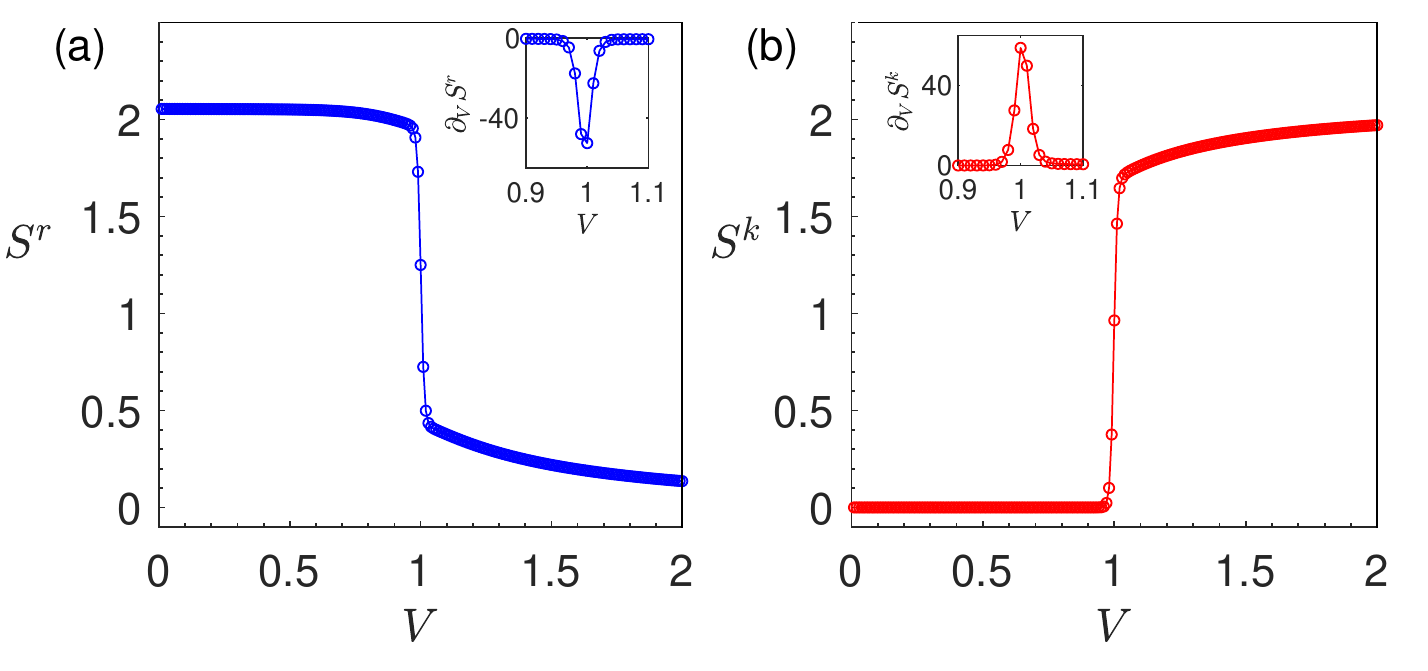}
\caption{{\bf MIT point from EE.} (a) and (b) respectively show the real-space and momentum-space EE and their derivative as a function of the potential strength $V$. Here, $J_L=0.5$.
}
\label{entropy}
\end{figure}

{\color{blue}\emph{Global phase diagram from entanglement.}}---In  non-Hermitian free fermionic systems, the left and right eigenvectors satisfy the bi-orthogonal condition. We need to provide the definition of density matrix in non-Hermitian systems~\cite{changEntanglement2020, herviouEntanglement2019, chenEntanglement2020,guoEntanglement2021} before moving forward. Below, we use the right and left operator $\psi_{R\alpha}^{\dag}$ and $\psi_{L\beta}^{\dag}$ to construct the right and left many-body states $|G_R\rangle$ and $|G_L\rangle$ of non-Hermitian system, where the real part of energy is regarded as filling level~\cite{ashidaNonHermitian2020, herviouEntanglement2019,changEntanglement2020}. Using these many-body states, the non-Hermitian density matrix can be expressed as $\rho=|G_R\rangle\langle G_L|$.  When partitioning the real space of system into two parts $A$ and $B$, and taking the partial trace over the part $B$, the reduced density matrix $\rho^{r}_{A}=\text{Tr}_{B} \rho$ ($r$ stands for real-space) can be used to compute EE $S^{r}=-\text{Tr}(\rho^{r}_{A}\ln \rho^{r}_A)$~\cite{PeschelCalculation2003}. By introducing the entanglement Hamiltonian $h^{E}$   via  $\rho^{r}_{A}=\exp(-h^E)$, the spectrum of entanglement Hamiltonian, i.e.,    ES  provides more information of the underlying system. For a non-Hermitian free fermionic system, the correlation matrix $C^r$, whose elements read $C^{r}_{mn}=\langle G_L| c_{m}^{\dag}c_{n}|G_R\rangle$, is related to entanglement Hamiltonian via $h^E=\log [(C^{r})^{-1}-\mathds{I}]$ with $\mathds{I}$ being the identity matrix. In addition, the eigenvalue $\epsilon^{r}_{i}$ of $h^E$ and the eigenvalue $\xi^{r}_{i}$ of $C^r$ is one to one correspondence $\epsilon^{r}_{i}=\log [(\xi^{r}_{i})^{-1}-1]$, so in the latter figures, we use $\xi^r_i$'s to denote ES.

The entanglement quantities, e.g., EE and ES, depend    on how Hilbert space is partitioned. In general, the real-space partition is the conventional choice to probe the localized states. At the same time, the momentum-space partition provides an insightful and complementary way to probe the extended states. When we change Fermi energy of the system, the states near Fermi energy have the most dominant contribution to ES\cite{lundgrenMomentumspace2019}. For the present purpose of exploring phase diagram of quasicrystals, we shall combine the two partitions. Specifically, the momentum partition divides the momentum space $(-\pi,\pi]$ into two parts $A$ and $B$, such as $A=(-\pi,0]$ and $B=(0,\pi]$. As for real-space partition, we choose nearly half partition $L_A=(L-1)/2$, where $L$ is odd. We use $S^{k}$ and $\xi^{k}_i$ to respectively denote EE and the $i$-th eigenvalue of ES when the momentum-space partition is adopted.

To study the entanglement properties of the model~(\ref{model1}), we numerically diagonalize the non-Hermitian Hamiltonian by varying potential $V$ with fixed $J_{L}=0.5$. As shown in Fig.~\ref{entropy},   we observe a phase transition point of the model~(\ref{model1}) located at $V_c=V=1$ where the derivative of EE diverges, meanwhile, the result is unchanged by changing the size of the subsystem $A$. This critical point, i.e., $V_c=1$, is a metal-insulator transition (MIT) point whose name will be much clearer in the calculation of ES. When we change the parameter $J_{L}$, the transition point is always invariant(detailed discussion see Supplemental Materials\cite{supmat}).  We will also prove that this point is analytically exact when $J_L=0$. In experiments,   the measurement of R\'enyi entropy has been studied~\cite{islamMeasuring2015}, which can be used to extract  the data of EE (i.e., von Neumann entropy)   in this work. Thus,  in Supplemental Material~\cite{supmat}, we also plot R\'enyi entropy   and find features similar to EE.

\begin{figure}[htbp]
\centering
\includegraphics[width=8.5cm]{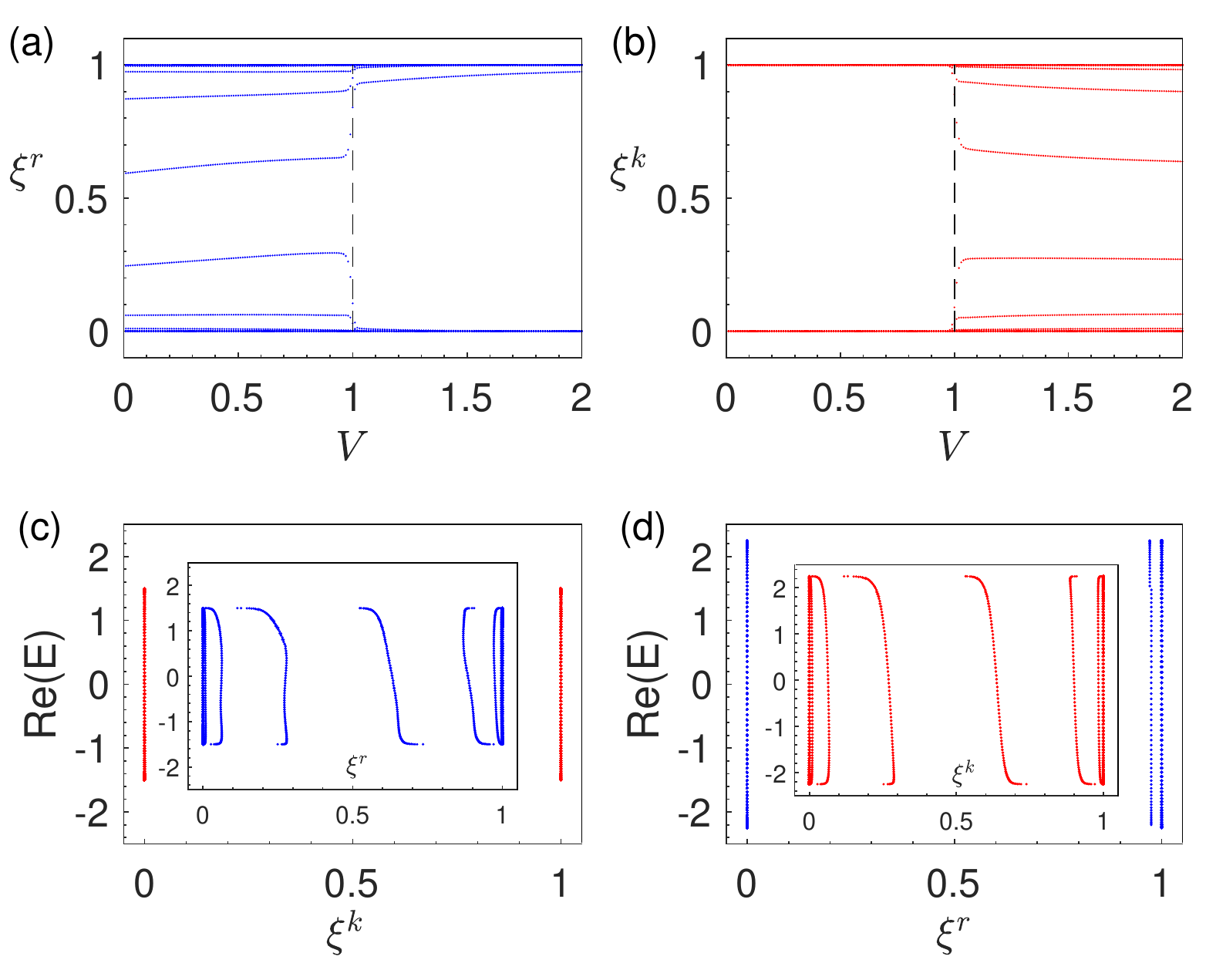}
\caption{{\bf Delocalization phase and localization phase from ES.} (a) and (b)  respectively show the real-space and momentum-space ES as a function of the potential strength. (c) shows the momentum-space ES as a function of real part of Fermi energy, and the inset is the real-space ES when $V=0.5$. (d) shows the real-space ES of real part of Fermi energy, and the inset is the momentum-space ES when $V=2$. Here, $J_{L}=0.5$.}\label{spectrum}
\end{figure}

Before discussing the ES of the model~(\ref{model1}), we consider the entanglement features of two kinds of typical quantum states, namely, plane wave state and completely localized state. Momentum-space ES of the plane wave state and real-space ES of completely localized state only have   $0$ and $1$ modes, and these modes do not contribute to EE at all. These special features can help us to distinguish different eigenstates in non-Hermitian quasicrystals (more details are available in Supplemental Material~\cite{supmat}). Keeping the above fact in mind, we investigate the real-space and momentum-space ES of   Eq.~(\ref{model1}). In Fig.~\ref{spectrum}(a) and (b), we demonstrate the real-space and momentum-space ES  by varying incommensurate potential $V$ and by taking account of nearly half partition and   half-filled occupation under PBC. We find all modes of real-space ES for the parameter range $V\in (1,2)$ and momentum-space ES for the parameter range $V\in (0,1)$ are very close to either $0$ or $1$. Therefore, combining these complementary features, we conclude that the occupied states in the parameter range $V\in (0,1)$ ($V\in (1,2)$) are physically extended states (physically localized states). Then the parameter range $V\in (0,1)$ ($V\in (1,2)$) represents a delocalization (localization) phase. Consequently, the phase transition of  the  model~(\ref{model1}) can be regarded as a delocalization-localization transition, justifying the name ``MIT point''  in the calculation of EE.

Next, we focus  two typical parameters $V=0.5$ and $2$ which respectively are located in  the delocalization and localization phases. As shown in Fig.~\ref{spectrum}(c) and (d), almost all eigenvalues of momentum-space (real-space) ES at $V=0.5$ ($V=2$) are located near $0$ or $1$, so we conclude that all eigenvalues have same properties. Furthermore, by varying the potential $V$, almost all eigenstates simultaneously are changed from extended states to localized states.  In conclusion, the mobility edge of the model (\ref{model1}) is absent.

\begin{figure}[htbp]
\centering
\includegraphics[width=8.5cm]{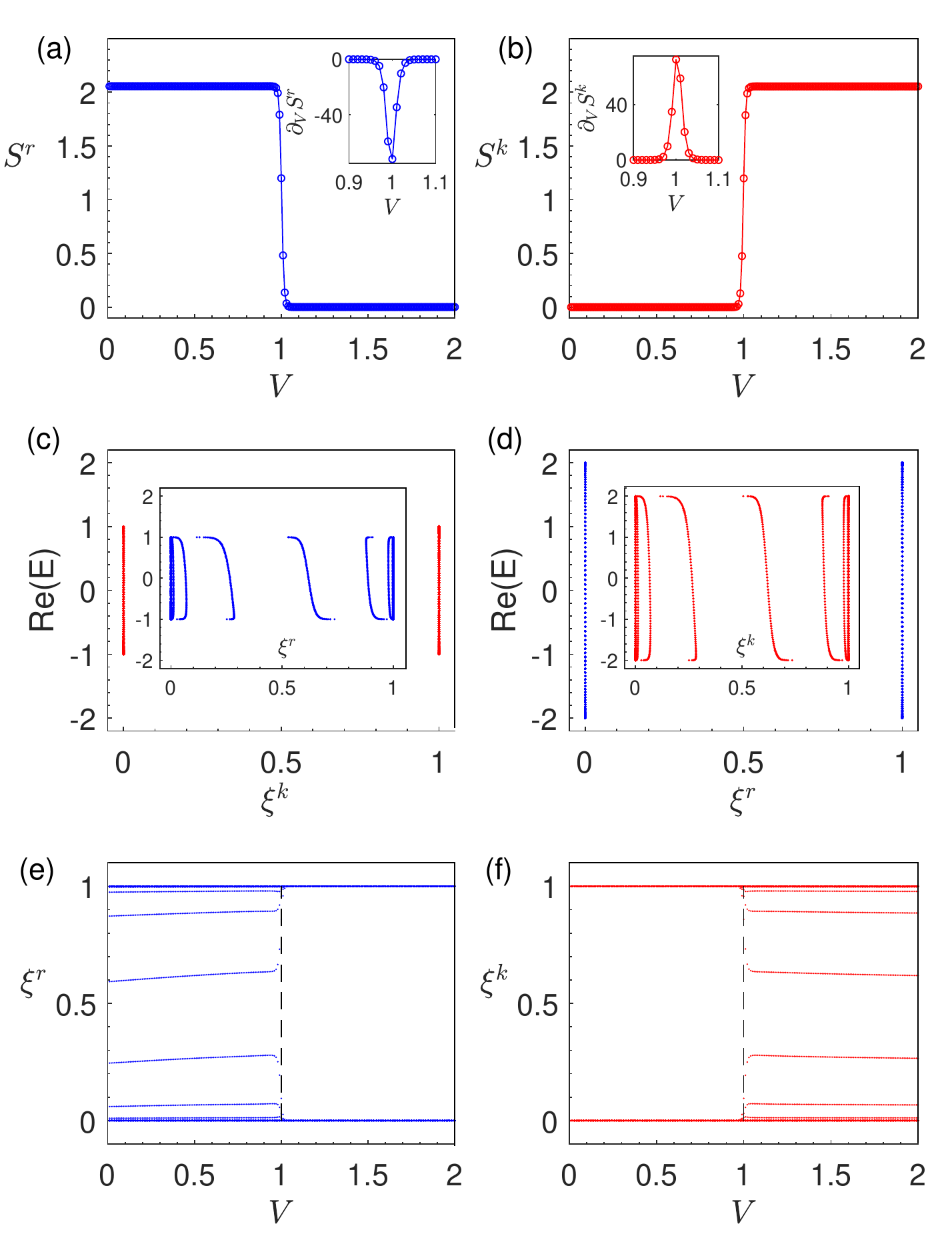}
\caption{{\bf Exactness of MIT point.} (a) and (b) are respectively   the real-space and momentum-space EE as a function of the potential strength. The two insets are the derivative of EE. (c) is the momentum-space ES as a function of real part of Fermi energy, and the inset is the real-space ES when $V=0.5$. (d) is the real-space ES of real part of Fermi energy, and the inset is the momentum ES when $V=2$. (e) and (f) respectively show the real-space and momentum-space ES as a function of the potential strength $V$. From the data in (e) and (f), one can numerically check that the validity of (\ref{eqn_duality}).  $J_L=0$ in all subfigures.}\label{NHentropy}
\end{figure}

{\color{blue}\emph{Analytically exact MIT point.}}---Below we shall show that the MIT point $V_c=1$ found in the numerical calculation of EE is analytically exact, when $J_{L}=0$ in the model~(\ref{model1}). Without loss of generality, we set $J_R=1$ again.

More precisely, when $J_L=0$, the ES eigenvalues of this model   with the same occupation and the same partition have the following nice identity between two ES eigenvalue sets:
\begin{align}
\left\{\xi^k_i(V)\right\}=\left\{\xi_i^r( {V}^{-1})\right\}\label{eqn_duality}\,.
\end{align}
Here $\left\{\xi^k_1(V),\xi^k_2(V),\xi^k_3(V),\cdots\right\}$ is the eigenvalue set of momentum-space ES when the potential strength is $V$. $\left\{ \xi^r_1(\frac{1}{V}),\xi^r_2(\frac{1}{V}),\xi^r_3(\frac{1}{V}),\cdots\right\}$)  is the eigenvalue set of real-space ES when the potential strength is $1/V$. The identity (\ref{eqn_duality}) means that the above two sets are identical.  Alternatively speaking, the identity (\ref{eqn_duality}) establishes an exact duality between  delocalization phase and localization phase parametrized by the potential parameter. In addition, $S^r(V)$ and $S^k(V^{-1})$ also have a dual relation as shown in Fig.~\ref{NHentropy}(a) and (b).

   In the previous discussion, we have introduced the significant entanglement feature in the delocalization (localization) phase: almost all eigenvalues of momentum-space (real-space) ES,   are very  close to either $0$ or $1$. At the same time, the real-space (momentum-space) ES of the delocalization (localization) phase doesn't possess such a   feature. If $V$ starts to increase near zero, the model evolves from a simple asymmetric hopping model and thus should be in the delocalization phase. If  $V$ starts to decrease near infinity,  the model should be in the localization phase.   Therefore,  one can directly verify that the identity (\ref{eqn_duality}) is consistent to the entanglement feature of delocalization and localization phases. Furthermore, the identity indicates that $V=V_c=1$ is very special. When the model is at this parameter point, the momentum-space ES and the real-space ES  of the same model are exactly the same, i.e.,
 \begin{align}
  \left\{\xi^k_i(1) \right\}=\left\{\xi^r_i(1)\right\}\,.
  \end{align}
In other words,  $V_c=1$ is a self-dual point where both real-space ES and momentum-space ES possess exactly the same behavior. It implies that this point belongs to neither delocalization phase nor localization phase, which must be the MIT point with delocalization-localization transition.

While a more mathematical proof is left to Supplemental Material~\cite{supmat}, below we present the most elementary ingredients towards the identity (\ref{eqn_duality}). One can start with the model with vanishing $J_L$.   Then the  model~(\ref{model1}) reduces to (model-I):
\begin{equation}
\begin{split}
\label{JLOR}
H^I_V=\sum_n\left(c_{n+1}^{\dag}c_{n}+V e^{-2\pi i \alpha n}c_n^{\dag}c_n\right) ~,
\end{split}
\end{equation}
where the subscript $V$ emphasizes the parameter-dependence. Introduce the Fourier transformation $\mathcal{F}$ and space inversion operator $\mathcal{P}$, where $\mathcal{P}$ transforms the lattice index $n$ to $-n$. By applying the two transformations on Eq.~(\ref{JLOR}), Eq.~(\ref{JLOR}) is then transformed to:
 $  \sum_{\tilde{m}}\left(V c_{\tilde{m}+1}^{\dag}c_{\tilde{m}}+     e^{-2\pi i \alpha \tilde{m}}c_{\tilde{m}}^{\dag}c_{\tilde{m}}\right )\,,
 $ where $\tilde{m}$ denotes momenta. If we interpret $\tilde{m}$ as lattice sites, then we immediately arrived at a new Hamiltonian (model-II):
\begin{equation}
\begin{split}
\label{model_2}
  {H}^{II}_V= \sum_{  n}\left(V c_{n+1}^{\dag}c_{ n}+     e^{-2\pi i \alpha  n}c_{n}^{\dag}c_{n}\right )\,.
\end{split}
\end{equation}
Interestingly, the expressions of model-I and model-II are related to each other by simply switching the coupling coefficients of the two Hamiltonian terms. Alternatively, $  {H}^{II}_V= V \sum_{  n}\left(  c_{n+1}^{\dag}c_{ n}+  V^{-1}   e^{-2\pi i \alpha  n}c_{n}^{\dag}c_{n}\right )= V H^{I}_{V^{-1}}$.
Since an overall numeric factor in Hamiltonian only scales energy spectrum but keeps entanglement quantities invariant,  we can conclude that   the real-space ES of model-II, denoted as a set $\{\xi_i^{II,r}(V)\}$ can be directly obtained by calculating the real-space ES of model-I when the potential is $V^{-1}$, i.e.,   $\left\{\xi^{II,r}_i(V)\right\}=\left\{\xi^{I,r}_i ( {V}^{-1})\right\}$. Furthermore, recalling the above construction of model-II from model-I where the spatial coordinates of model-II  originally come from the momenta of model-I, we  have another identity: $\left\{\xi^{II,r}_i(V)\right\}=\left\{\xi^{I,k}_i ( {V})\right\}$. Combining all facts together and removing $I$, we end up with  the identity (\ref{eqn_duality}).

%
%
%
%
%
%

From the numerical aspect, we use EE and ES to display the identical relation~(\ref{eqn_duality}) of the model~(\ref{model1}) with $J_{L}=0$. As shown in Fig.~\ref{NHentropy} (a) and (b), real-space and momentum-space EE are symmetric with respect to $V=1$ and the self-duality point $V=1$ is the transition point. Meanwhile, we choose the parameters $V=0.5$ and $2$ to study the ES of the model~(\ref{model1}) to demonstrate the identity~(\ref{eqn_duality}). As shown in Fig.~\ref{NHentropy} (c) and (d), when we rescale the energy of the Hamiltonian~(\ref{model1}) at $V=2$ to $\frac{1}{2}$ of original, the real (momentum) space ES of Hamiltonian~(\ref{model1}) at $V=0.5$ is identical with momentum (real) space ES at $V=2$. Finally, we consider the ES of the model~(\ref{model1}) as a function of potential $V$, these data also show the real and momentum space ES of the model~(\ref{model1}) are symmetric with respect to $V=1$ in Fig.~\ref{NHentropy} (e) and (f), therefore, we verify the duality of Hamiltonian~(\ref{model1}) with $J_{L}=0$ again.

\begin{figure}[htbp]
\centering
\includegraphics[width=8.5cm]{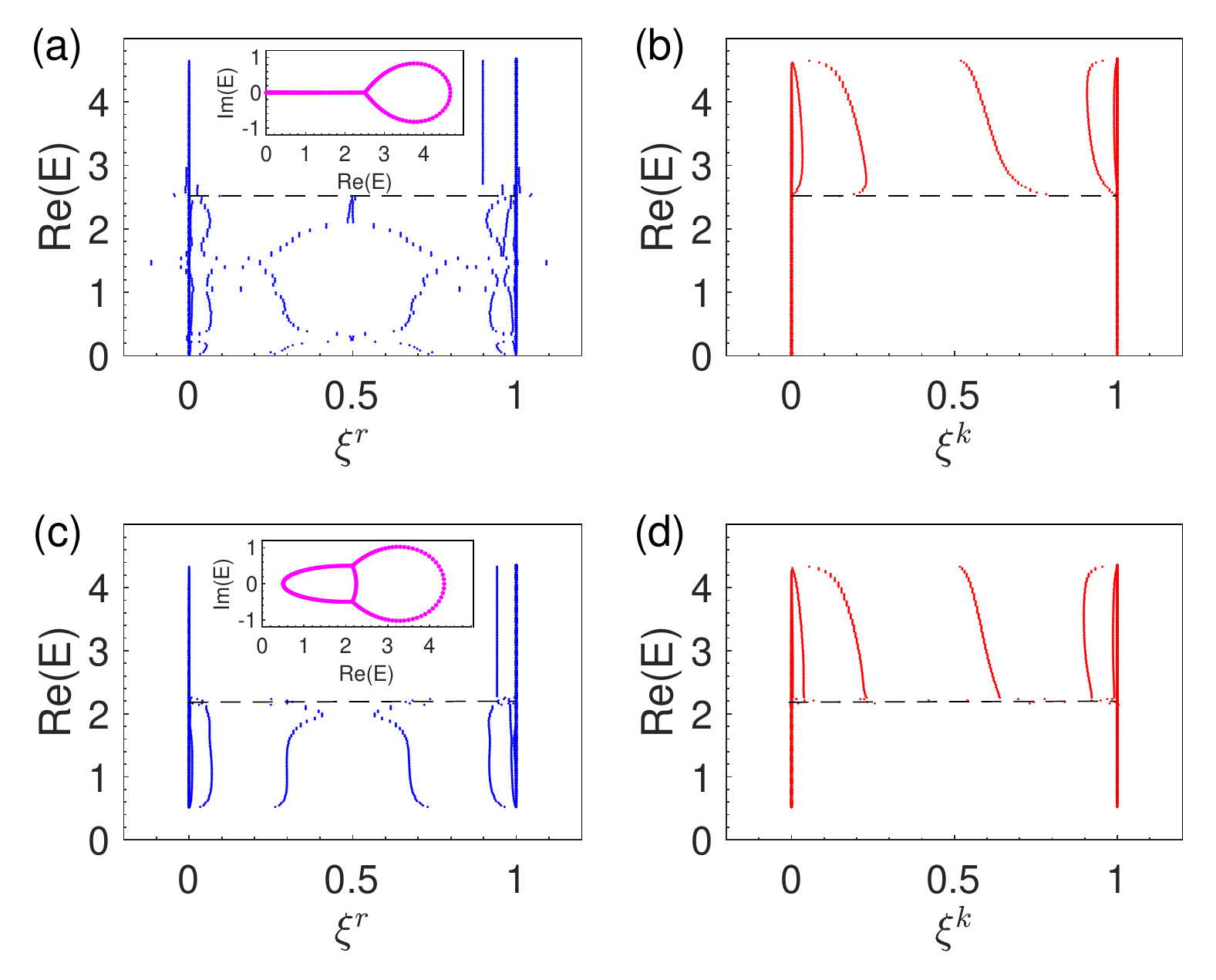}
\caption{{\bf Mobility edge from ES.}  (a) and (b) are respectively  the real-space and momentum-space ES of the model~(\ref{eq:me_model}) with  symmetric hopping $J_{R}=J_{L}=1$ by varying the real part of Fermi energy.   (c) and (d) are respectively the real-space and momentum-space ES of the model~(\ref{eq:me_model}) with asymmetric hopping $J_{R}=1, J_{L}=0.5$. Both insets in (a) and (c) are the energy spectrum plotted in complex plane.}\label{mobility}
\end{figure}

{\color{blue}\emph{Mobility edge  in general models from entanglement.}}---In the above discussion, we have found that  the  model~(\ref{model1}) doesn't have mobility edge. Usually, mobility edge in non-Hermitian quasicrystals can be realized by long-range hopping or special on-site potential. Thus, we introduce the  model below and identify mobility edge from the perspective of entanglement:
\begin{equation}
\begin{split}
\!\!\!\!H\!=\!\sum_{n}(J_Rc_{n+1}^{\dag}c_n\!+\!J_Lc_{n}^{\dag}c_{n+1})\!+\!\sum_n \!\frac{V}{1-a e^{i 2 \pi \alpha n}}c_{n}^{\dag}c_{n},\!
\end{split}
\label{eq:me_model}
\end{equation}
where $\alpha$ is the incommensurate ratio and we fix $\alpha=\sqrt{2}\approx \frac{239}{169}$ with lattice size $L=169$, $a=0.5$ and $V=2$. As shown in Fig.~\ref{mobility}(a) and (b) where we consider symmetric hopping and (\ref{eq:me_model}) reduces to the model in Ref.~\cite{liuGeneralized2020}, real-space and momentum-space ES have distinctive features near the mobility edge (marked by the horizontal dashed lines). There, the extended states suddenly become the localization states. As a result, mobility edge is identified from entanglement, where the real (complex) energy spectrum [i.e., the inset in Fig.~(\ref{mobility})(a)] corresponds to extended (localized) states.

On the other hand, as shown  in Fig.~(\ref{mobility})(c) and (d) where asymmetric hopping is considered, the data of ES near the mobility edge also exhibit distinctive features. But now the energy of extended states are  complex and the relation between real (complex) energy and the extended (localized) states is absent, as shown in the inset of Fig.~(\ref{mobility})(c).

In addition to the above two models, we can further generalize the model~(\ref{model1}) into a more general form:
\begin{equation}
\begin{split}
 H=&\sum_{n,i,j}(J_Rc_{n+i}^{\dag}c_n+J_Lc_{n}^{\dag}c_{n+j})\\
&+\sum_n(V_Re^{2\pi i \alpha n}+V_Le^{-2\pi i \alpha n})c_n^{\dag}c_n ~,
\end{split}
\label{eq:gen_NH_AAH}
\end{equation}
where $n$ is the index for lattice site. When   $J_R=J_L,V_R=V_L$, the model~(\ref{eq:gen_NH_AAH}) reduces  to the famous   AAH model~\cite{aubry1980analyticity}; when $J_R\neq J_L,V_R=V_L$, the model~(\ref{eq:gen_NH_AAH}) reduces to the model studied in Ref.~\cite{longhiPhase2021}; when $J_R=J_L,V_R=0$, the model~(\ref{eq:gen_NH_AAH})  reduces to the model studied in Ref.~\cite{longhiMetalinsulator2019}. Besides, when we set  $J_R,J_L\in\mathds{C}$, it is equivalent to implement external magnetic flux in the model~(\ref{eq:gen_NH_AAH}).
For our model~(\ref{model1}) where $J_{R}\neq J_{L}, V_{R}\neq V_{L}$, the existence of two distinct sources of non-Hermiticity plays a critical role in guaranteeing the identity (\ref{eqn_duality}).

{\color{blue}\emph{Discussion and outlook.}}---In this work, we introduced the powerful entanglement approach to unveil exotic non-Hermitian quantum effect in   non-Hermitian quasicrystal chains where   asymmetric hopping and complex potential coexist. We obtained the global phase diagram from both numerical and analytical analysis of entanglement. The MIT point was proved to be exact. We also studied mobility edge by means of entanglement in general models.

In our study of the model~(\ref{model1}), we found the real parts of all eigenvalues of correlation matrix are located in the interval $[0,1]$ similar to the Hermitian system. Nevertheless, we found ES of the model~(\ref{eq:me_model}) with symmetric hopping includes anomalous eigenvalues whose real parts are outside the interval $[0,1]$, as shown in Fig.~(\ref{mobility})(a). As shown in Fig.~\ref{mobility}(c) and (d), ES of the model~(\ref{eq:me_model}) with asymmetric hopping does not have anomalous value, and its energy spectrum doesn't have real-complex transition.      Recently, the phenomenon of   anomalous values of ES at critical point with exceptional point has been demonstrated and discussed in various    non-Hermitian crystals~\cite{changEntanglement2020,lee2020exceptional}. In our non-Hermitian quasicrystal systems, we infer that the anomalous values of ES  may have an intrinsic tie to the exceptional point of energy spectrum, which is left to future study. In addition, we may also ask the following questions:  How can we apply entanglement to characterize non-Hermitian random disorder models?  How can we obtain more information from entanglement Hamiltonian, such as localization length of localized states?

We thank the following funding sources: Guangdong Basic and Applied Basic Research Foundation under Grant No.~2020B1515120100, NSFC Grant (No.~11847608 \& No.~12074438).

\appendix

\section{More discussion about the model~(1) in the main text}\label{sec:appendix}

\subsection{Energy spectrum of model~(1) in the main text}
\begin{figure}[htbp]
\centering
\includegraphics[width=8.5cm]{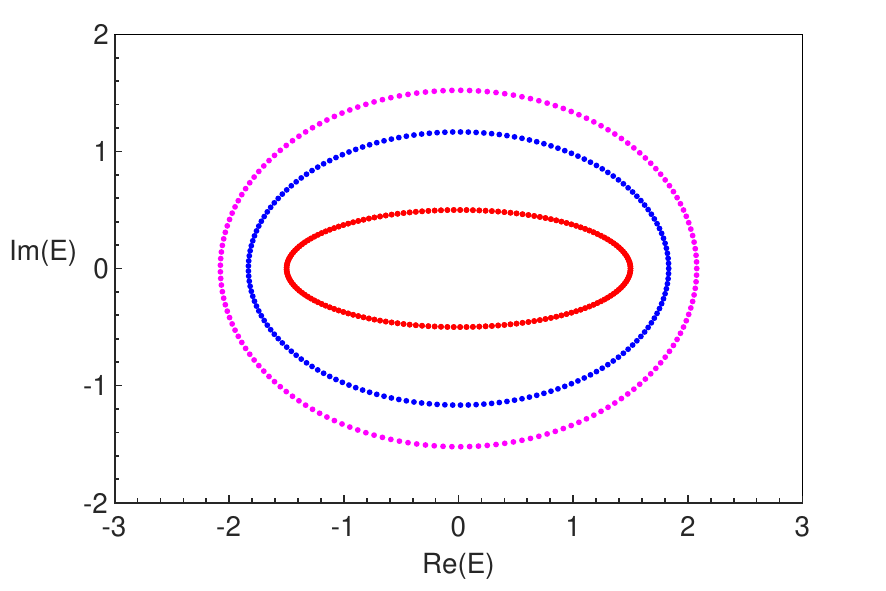}
\caption{Energy spectrum of non-Hermitian quasicrystal model with $V=0.5(\text{red}),0.8(\text{red}),1.5(\text{blue}),1.8(\text{purple})$ in complex plane.  Here, $J_L=0.5$. }\label{energy}
\end{figure}

In this section, we give a detailed analysis about the energy spectrum of model~(1) in some parameter region. Under periodic boundary condition (PBC), the energy spectra of model~(1) in real and momentum space are identical under the thermodynamics limit $L\rightarrow \infty$. We use the eigenvalue equation to analyse the eigenenergy of original Hamiltonian, written as
\begin{equation}
E u_{n}=J_Ru_{n+1}+J_Lu_{n-1}+V e^{- 2 \pi i \alpha n} u_{n}\ ,
\end{equation}
where $n$ is the index for lattice site. We introduce a Fourier transformation for $u_n$,
\begin{equation}
u_{n}=\frac{1}{\sqrt{L}} \sum_{k} u_{k} e^{-i k n} \ , \quad u_{k}=\frac{1}{\sqrt{L}} \sum_{n} u_{n} e^{i k n} \ ,
\end{equation}
where $k=2 m \pi \alpha\mod 2\pi,\ m\in \mathbb{Z}$, then
\begin{equation}
Eu_m=Vu_{m-1}+J_R e^{-i2\pi \alpha m}u_m+J_Le^{i2\pi\alpha m}u_m \ ,
\end{equation}
where $m$ denotes momenta. For a finite-size system, if we take periodic boundary condition(PBC), we need to approximate the irrational $\alpha$ by $\alpha=\frac{M}{L}$ with $M$ and $L$ being coprimes. The momentum should be relabeled by $k=2 m \pi \frac{M}{L}\mod 2\pi$ with $m=1,2,...,L$. From Ref.~\cite{longhiMetalinsulator2019}, the energy spectrum can be obtained from the roots $E_n$ of the following characteristic polynomial $P(E)$ of order $L$,
\begin{equation}
P(E)=\prod_{m=1}^L(E-W_m)-V^L
\end{equation}
with $W_m=J_R e^{-i2\pi \alpha m}+J_Le^{i2\pi\alpha m}$. Combined with the numerical result, the roots of polynomial $P(E)$ is $E_{m_0}=J_R e^{-i2\pi \alpha m_0}+J_Le^{i2\pi\alpha m_0}$ for $V<J_R$ in the large $L$ limit. Furthermore, with increasing potential $V$ until $V=J_R$, its energy spectrum is invariant as shown in Fig.(\ref{energy}). When $V>J_R$, the energy spectrum is a larger ellipse in complex energy plane\cite{longhiMetalinsulator2019}.

\subsection{Eigenstates of the model~(1) in the main text}
\begin{figure}[htbp]
\centering
\includegraphics[width=8.5cm]{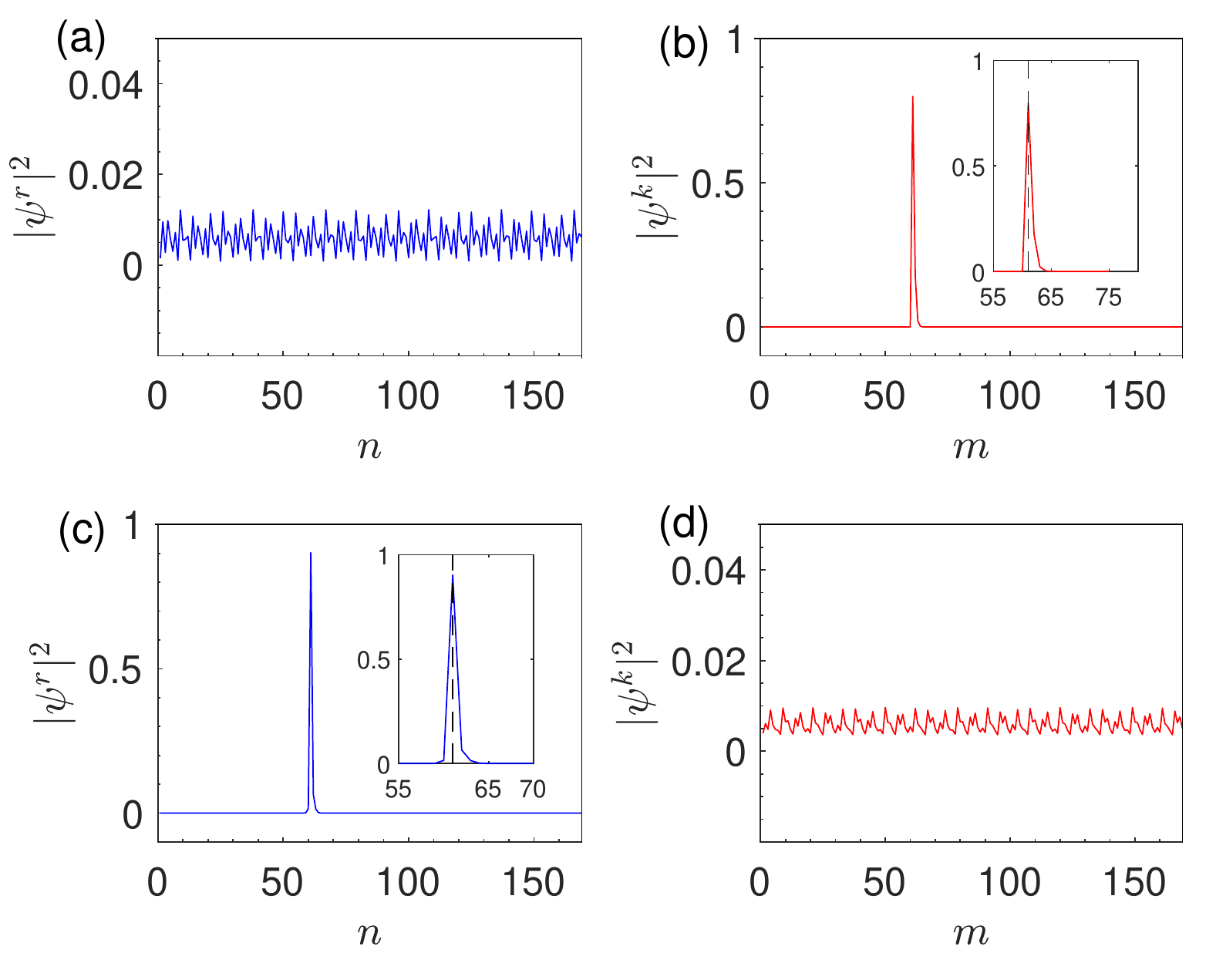}
\caption{(a) and (b) are the probability distribution of the typical eigenvector of Hamiltonian in real space and momentum-space, respectively, when $J_L=0.5,V=0.5$ under PBC. (c) and (d) are the probability distribution of the typical eigenvector of Hamiltonian in real space and momentum space, respectively, when $J_L=0.5,V=2$. }\label{state1}
\end{figure}

The dual features of non-Hermitian AAH model can also be exhibited in the eigenstates. In other words, the nonreciprocal hopping and incommensurate complex potential can be transformed into each other. We choose two parameters $V=0.5$ and $V=2$ to display the dual characteristic of eigenstates. As shown in Fig.\ref{state1}(b), the momentum-space localized eigenstates of delocalization phase have the asymmetrical left and right localization length. After we do fourier transformation for the model~(1), the complex incommensurate potential of Hamiltonian~(1) in real space would be transformed into nonreciprocal hopping term of Hamiltonian~(1) in momentum space. Hence, the unequal localization length in momentum space stems from complex incommensurate potential. As illustrated in Fig.(\ref{state1})(c), the physical localized states also have the asymmetrical localization length, and this feature is directly coming from the nonreciprocal hopping. Ref.~\cite{kawabataNonunitary2021} shows that the nonreciprocal hopping models have two types transfer matrices leading the asymmetrical localization length.

\subsection{Imaginary part of ES}
\begin{figure}
\centering
\includegraphics[width=8.5cm]{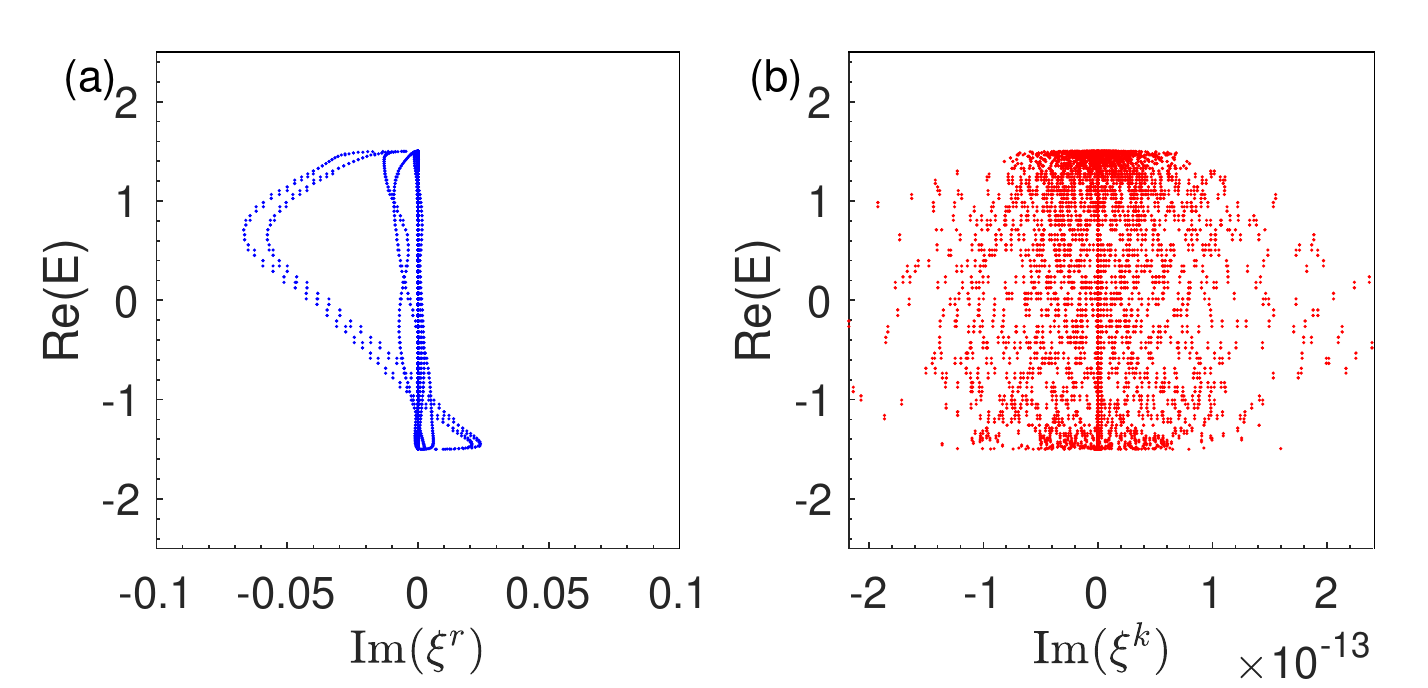}
\caption{(a) is imaginary part of the real-space ES as a function of the real part of Fermi energy. (b) is imaginary part of the momentum-space ES as a function of the real part of Fermi energy, when $J_L=0.5,V=0.5$. }\label{imaginary1}
\end{figure}

\begin{figure}
\centering
\includegraphics[width=8.5cm]{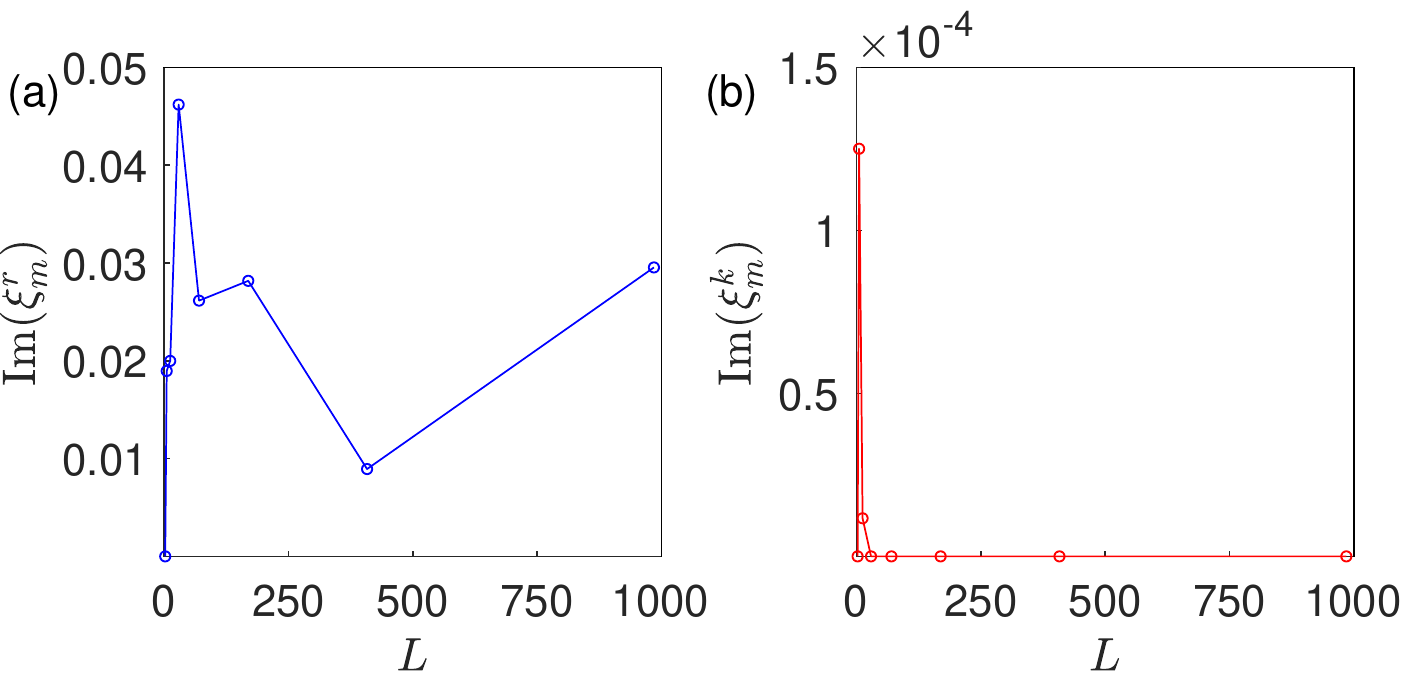}
\caption{(a) and (b) are the maximal imaginary part of real-space and momentum-space ES with the increase of lattice size and half-partition, respectively, when $J_L=0.5,V=0.5$ and $L=2,5,12,29,70,169,408,985$. }\label{imaginary2}
\end{figure}

In non-Hermitian quasicrystal models, the real-space ES would have some finite imaginary value as shown in Fig.(\ref{imaginary1}). However, the imaginary part of ES are much smaller than the real part and very close to 0. The imaginary part of momentum-space ES can be ignored as shown in Fig.(\ref{imaginary1})(b). In addition, with the increase of lattice size, the imaginary part of ES become smaller and smaller, as shown in Fig.(\ref{imaginary2}). Then, we conjecture that the finite imaginary part originate from the size effect and calculation error. Meanwhile, the imaginary part does not influence our study for model~(1), then we ignore the imaginary part and use the real part of ES to display the properties of model~(1).

\subsection{R\'enyi entropy}
\begin{figure}[htbp]
\centering
\includegraphics[width=8.5cm]{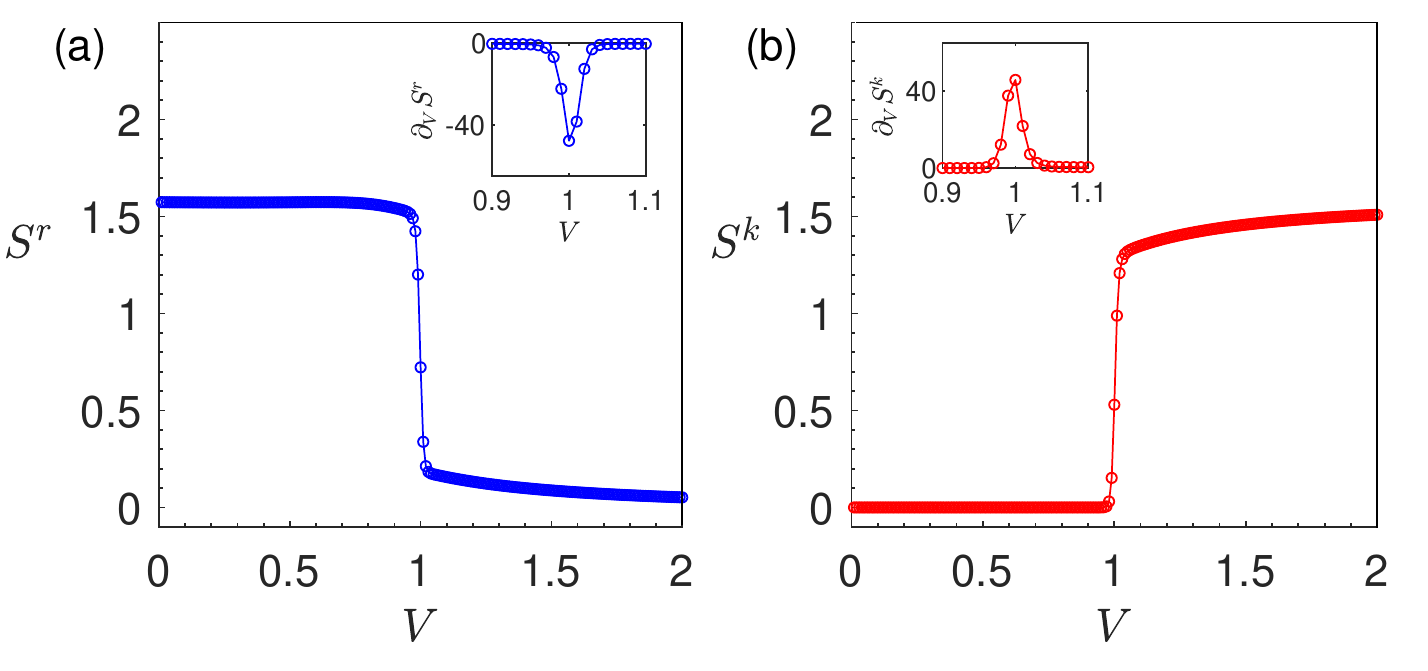}
\caption{\textbf{MIT transition point from R\'enyi entropy.} (a) and (b) respectively show the real-space and momentum-space R\'enyi entropy and their partial derivative as a function of the potential $V$ when $J_L=0.5$. }\label{renyientropy}
\end{figure}

The definition of R\'enyi entropy is
\begin{equation}
S^n=\frac{1}{1-n}\ln(\text{Tr}  \rho_A^n) \,.
\end{equation}
Here we calculate the R\'enyi entropy to demonstrate the transition properties of model~(1). As shown in Fig.(\ref{renyientropy}), $S^2$ also shows that the potential $V=1$ is corresponding to the transition point.

\subsection{The transition point of the model Eq.(1) with $J_L<J_R$ is invariant.}
\begin{figure}
\centering
\includegraphics[width=8.5cm]{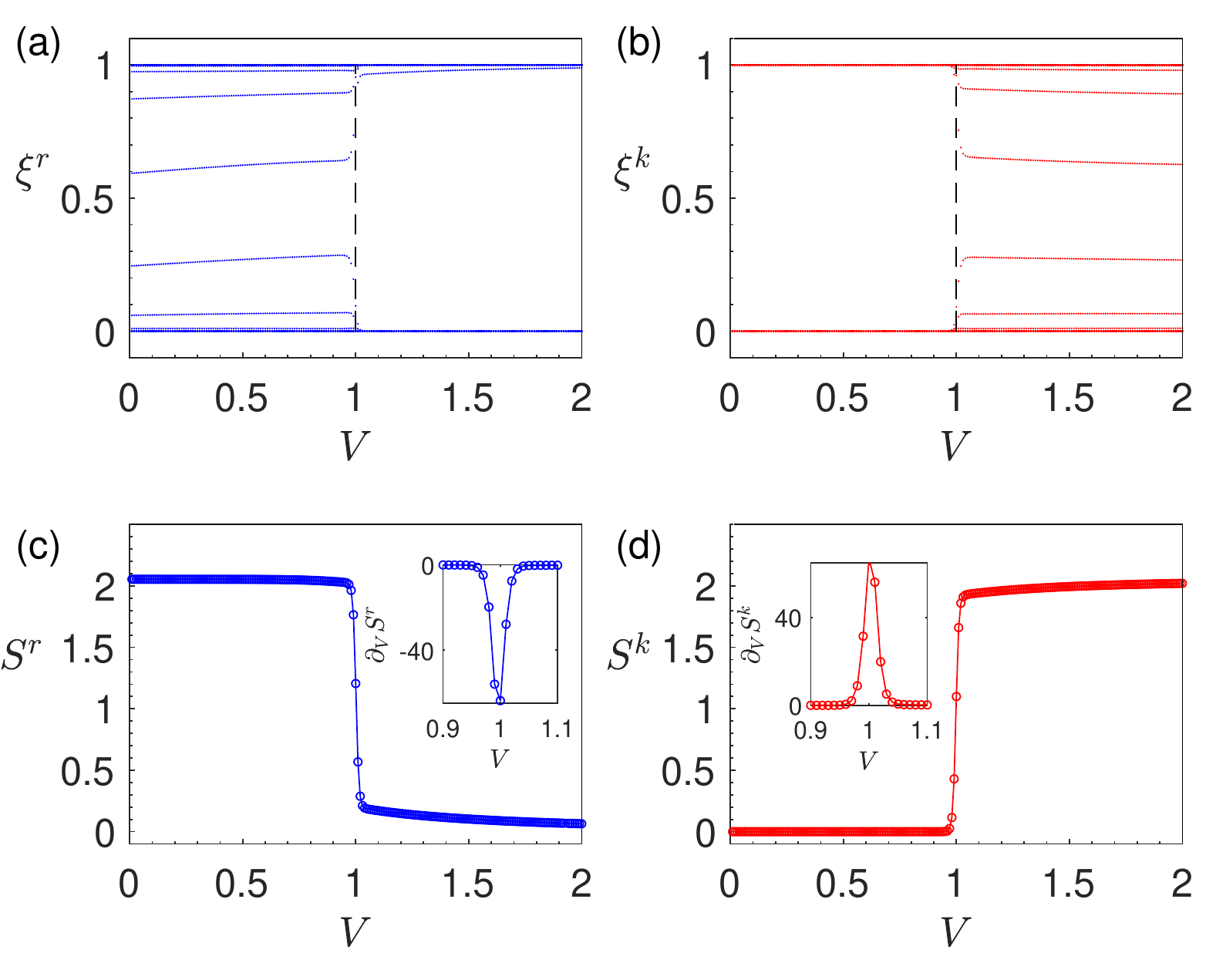}
\caption{(a) and (b)  respectively show the real-space and momentum-space ES as a function of the potential strength. (c) and (d) respectively show the real-space and momentum-space EE and their derivative as a function of the potential strength $V$. Here, $J_L=0.2$.}\label{entangle02}
\end{figure}

\begin{figure}
\centering
\includegraphics[width=8.5cm]{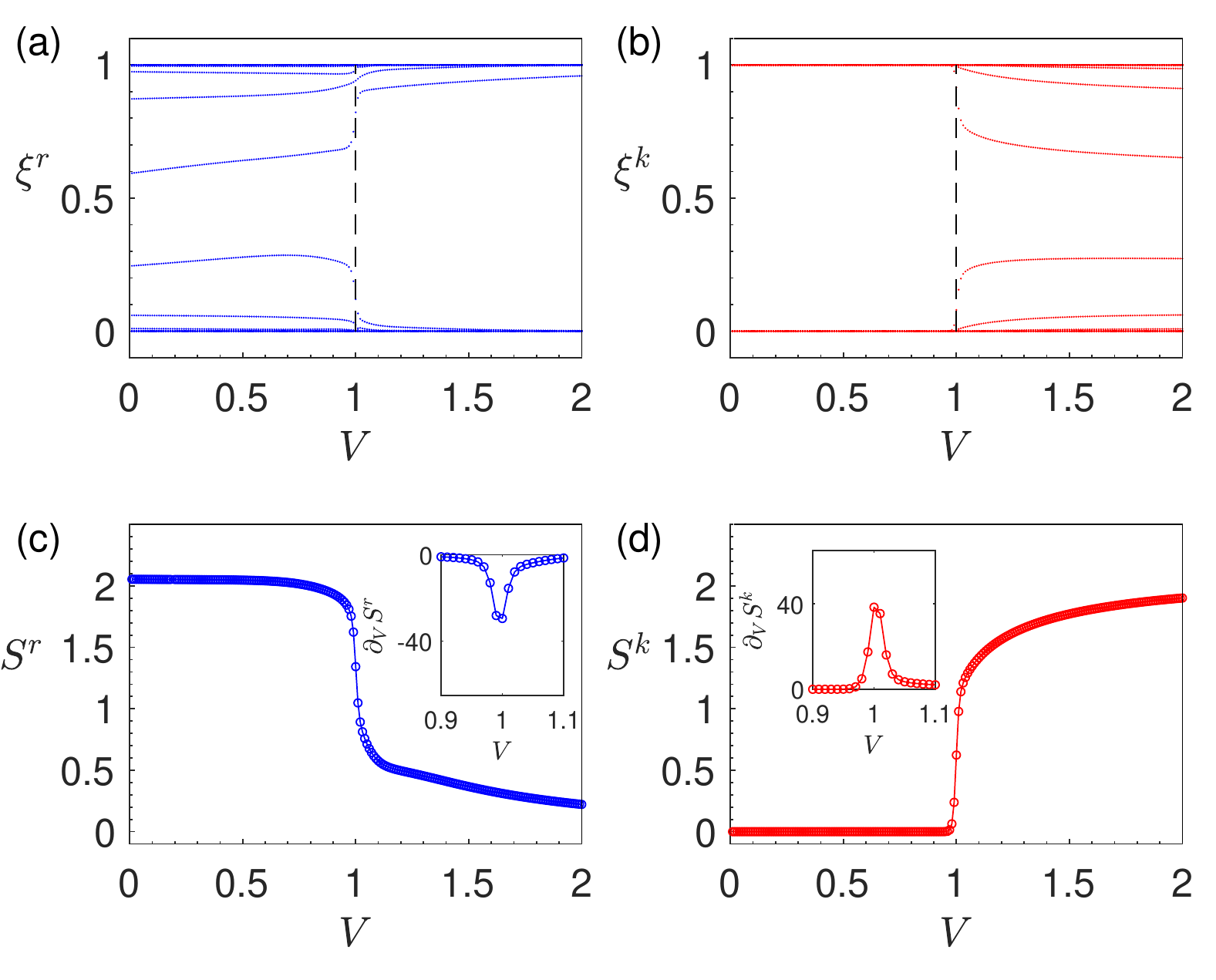}
\caption{(a) and (b)  respectively show the real-space and momentum-space ES as a function of the potential strength. (c) and (d) respectively show the real-space and momentum-space EE and their derivative as a function of the potential strength $V$. Here, $J_L=0.9$.}\label{entangle09}
\end{figure}

In the main text, we have exactly obtained the transition point when $J_L=0$, but we cannot analytically obtain the transition point when $J_L$ is nonzero. By using numerical simulation, we can still find the transition point with $J_L<J_R$ is unchanged along varying $J_L$. To be specific, we choose two parameter points $J_L=0.2$ and $J_L=0.9$ to calculate the entanglement spectrum and entropy. As shown in Fig.(\ref{entangle02}) and Fig.(\ref{entangle09}), the entanglement spectrum and entropy clearly demonstrate the transition point $J_R=V=1$, then to simplify the discussion in the main text, we choose a typical point $J_{L}=0.5$ to show the numerical result of the model Eq.(1).

\section{Entanglement in non-Hermitian systems.}
In the non-Hermitian systems, the left and right eigenvectors satisfy the biorthogonal condition. For a non-Hermitian free fermionic Hamiltonian $H=\sum_{mn}c^{\dag}_{m}\mathcal{H}_{mn}c_{n}$, where $\mathcal{H} \neq \mathcal{H}^{\dag}$, and $c^{\dag}_{m},c_{n}$ are fermionic operators and satisfy $\{c^{\dag}_{m},c_{n}\}=\delta_{mn}$. The diagonalizable $\mathcal{H}$ have a complete set of biorthogonal eigenvectors $\{|R_{\alpha}\rangle,|L_{\beta}\rangle\}$\cite{Brody_2013}, constructed by $\mathcal{H}|R_{\alpha}\rangle=\epsilon_{\alpha}|R_{\alpha}\rangle$ and $\mathcal{H}^{\dag}|L_{\beta}\rangle=\epsilon_{\beta}^*|L_{\beta}\rangle$, where $|R_{\alpha}\rangle$ is right eigenvector and $|L_{\beta}\rangle$ is left eigenvector. They satisfy $\langle L_{\beta}|R_{\alpha} \rangle=\delta_{\alpha\beta}$ and $\sum_{\alpha}|R_{\alpha} \rangle\langle L_{\alpha}|=\mathds{I}$, where $\mathds{I}$ is the identity matrix. Therefore, the Hamiltonian can be written as:
\begin{equation}
\mathcal{H}=\sum_{\alpha} \epsilon_{\alpha}|R_{\alpha} \rangle\langle L_{\alpha}| \ .
\end{equation}

From the left and right eigenvectors, the left and right creation operators can be introduced as $\psi_{R\alpha}^{\dag}|0\rangle\equiv|R_{\alpha} \rangle, \psi_{L\beta}^{\dag}|0\rangle\equiv|L_{\beta} \rangle$, and satisfy $\{\psi_{L\alpha},\psi_{R\beta}^{\dag}\}=\delta_{\alpha\beta}$. Different from the normal fermionic operator, it is named as bi-fermionic operator. We can use the bi-fermionic operator to construct left and right many-body states of non-Hermitian system,

\begin{equation}
|G_R\rangle=\prod_{\alpha \in \text{occ.}} \psi_{R\alpha}^{\dag}|0\rangle \ , |G_L\rangle=\prod_{\beta \in \text{occ.}}\psi_{L\beta}^{\dag}|0\rangle \ ,
\end{equation}
where $\text{occ.}$ denotes the occupied states. So we can use these ground states to obtain the reduced density matrix with partition.


Next, we discuss the partition of different degree of freedom of Hilbert space. The real-space partition divides the real-space chain into two parts $A$ and $B$, and the momentum-space partition divides the momentum space $(-\pi,\pi]$ into two parts $A$ and $B$, such as $A=(-\pi,0]$ and $B=(0,\pi]$. For a translationally invariant system, the wave vectors $k\in(-\pi,0]$ and $k\in (0,\pi]$ represent the forward and backward traveling states, respectively. If adding random disorder into this system, we can use momentum-space partition to probe the entanglement between the right movers and left movers\cite{mondragonshemCharacterizing2013,mondragonshemSignatures2014}.

We consider the entanglement features of two kinds of typical quantum states, namely, plane wave state and completely localized state. The plane wave state is given as $|k\rangle=c_k^{\dag}|0\rangle$, where $k$ is the wave vector. Then, the corresponding real-space correlation matrix and momentum-space correlation matrix read
\begin{equation}
\begin{split}
C_{r_mr_n}&=\langle k|c^{\dag}_{r_m}c_{r_n}|k\rangle=\frac{e^{-ik(r_m-r_n)}}{N} \ ,\\
C_{k_mk_n}&=\langle k|c^{\dag}_{k_m}c_{k_n}|k\rangle=\delta_{k_m k}\delta_{k_n k} \ ,
\end{split}
\end{equation}
where we restrict the indices $r_m,r_n\in[1,\frac{N}{2}], \ k_m,k_n\in(-\pi,0]$. We find that there are some eigenvalues of real-space correlation matrix located in the range $[0,1]$ and momentum-space correlation matrix only have $0$ or $1$ eigenvalues. Therefore, we can directly estimate the physical extended state from momentum-space correlation matrix. For a completely localized state $|r\rangle =c^{\dag}_r|0\rangle$, and its real-space correlation matrix and momentum-space correlation matrix read
\begin{equation}
\begin{split}
C_{r_mr_n}&=\langle r|c^{\dag}_{r_m}c_{r_n}|r\rangle=\delta_{r_m r}\delta_{r_n r} \ ,\\
C_{k_mk_n}&=\langle r|c^{\dag}_{k_m}c_{k_n}|r\rangle=\frac{e^{-ir(k_m-k_n)}}{N} \ .
\end{split}
\end{equation}
It is obvious that we can estimate the physical localized state from real-space correlation matrix.

\section{Duality}
In this part, we prove the duality in AAH model in detail. For a general AAH model, the eigenvalue equation is written as,
\begin{equation}
\!E^I u_{n}\!\!=\!\!J_Ru_{n+1}+\!\!J_Lu_{n-1}+\!\!V_R e^{i2 \pi \alpha n}u_{n}+\!\!V_Le^{-i2 \pi \alpha n}  u_{n}\!\! \ .
\label{eq:gen_AAH}
\end{equation}
The Fourier transformation of $u_n$,
\begin{equation}
u_{n}=\frac{1}{\sqrt{L}} \sum_{k} u_{k} e^{-i k n}, \quad u_{k}=\frac{1}{\sqrt{L}} \sum_{n} u_{n} e^{i k n} \ ,
\end{equation}
then
\begin{equation}
\begin{split}
&E^I \frac{1}{\sqrt{L}} \sum_{k} u_{k} e^{-i k n}=J_R\frac{1}{\sqrt{L}} \sum_{k} u_{k} e^{-i k (n+1)}+\\
&J_L\frac{1}{\sqrt{L}} \sum_{k} u_{k} e^{-i k (n-1)}+V_Re^{i2 \pi \alpha n} \frac{1}{\sqrt{L}} \sum_{k} u_{k} \\
&e^{-i k n}+V_Le^{-i2 \pi \alpha n}\frac{1}{\sqrt{L}} \sum_{k} u_{k} e^{-i k n}\\
&=\frac{1}{\sqrt{L}} \sum_{k} u_{k}e^{-i k n}(J_R e^{-ik}+J_Le^{ik})+\\
& \frac{1}{\sqrt{L}} \sum_{k} (V_Ru_{k}e^{-i (k -2\pi \alpha)n}+V_L u_{k}e^{-i(k+2\pi \alpha)n})\\
&=\frac{1}{\sqrt{L}} \sum_{k} u_{k}e^{-i k n}(J_R e^{-ik}+J_Le^{ik})+\\
&\frac{1}{\sqrt{L}} \sum_{k} (V_R u_{k+2\pi \alpha}+V_Lu_{k-2\pi \alpha})e^{-ikn} \ ,
\end{split}
\end{equation}
then
\begin{equation}
\begin{split}
E^Iu_k=J_R e^{-ik}u_k+J_Le^{ik}u_k+V_R u_{k+2\pi \alpha}+V_Lu_{k-2\pi \alpha} \ .
\end{split}
\end{equation}
where $k=2 m \pi \alpha\mod 2\pi,\ m\in \mathbb{Z}$. We can rewrite the equation,
\begin{equation}
\begin{split}
\!\!E^{II}u_m\!\!=\!(\!\!J_R e^{-i2\pi \alpha m}+\!\!J_Le^{i2\pi \alpha m})u_m+\!\!V_R u_{m+1}+\!\!V_Lu_{m-1}\! \ ,
\label{eq:gen_mom_eigen}
\end{split}
\end{equation}
where m is the index for the quasi-momenta with the interval being $2\pi \alpha$. For a finite-size system, if we take periodic boundary condition(PBC), we need to approximate the irrational $\alpha$ by $\alpha=\frac{M}{L}$ with $M$ and $L$ being coprimes. The momentum can be relabeled by $k=2 m \pi \frac{M}{L}\mod 2\pi$ with $m=1,2,...,L$.

When $J_L=0$ and $V_R=0$, then the eigenvalue equation~(\ref{eq:gen_AAH}) become
\begin{equation}
E^{I}u_m=J_Ru_{m+1}+V_L e^{-i2\pi \alpha m}u_m \, .\label{equ_eigen2}
\end{equation}
and the eigenvalue equation~(\ref{eq:gen_mom_eigen}) become
\begin{equation}
E^{II}u_m=V_Lu_{m-1}+J_R e^{-i2\pi \alpha m}u_m \, .\label{equ_eigen1}
\end{equation}
Next, we introduce a space inversion operator $\mathcal{P}$ acting on it, so the Eq.~(\ref{equ_eigen1}) is transformed to
\begin{equation}
E^{II}u_m=V_Lu_{m+1}+J_R e^{-i2\pi \alpha m}u_m \, .\label{equation2}
\end{equation}
When we interpret the $m$ as lattice sites, then we obtain a new model~(\ref{equation2}).
If we set $J_{R}=V_{L}$, Eq.~(\ref{equation2}) is equivalent to Eq.~(\ref{equ_eigen2}). Therefore, the two models~(\ref{equ_eigen2}) and (\ref{equation2}) are mutually duality.

\begin{figure}[htbp]
\centering
\includegraphics[width=8.5cm]{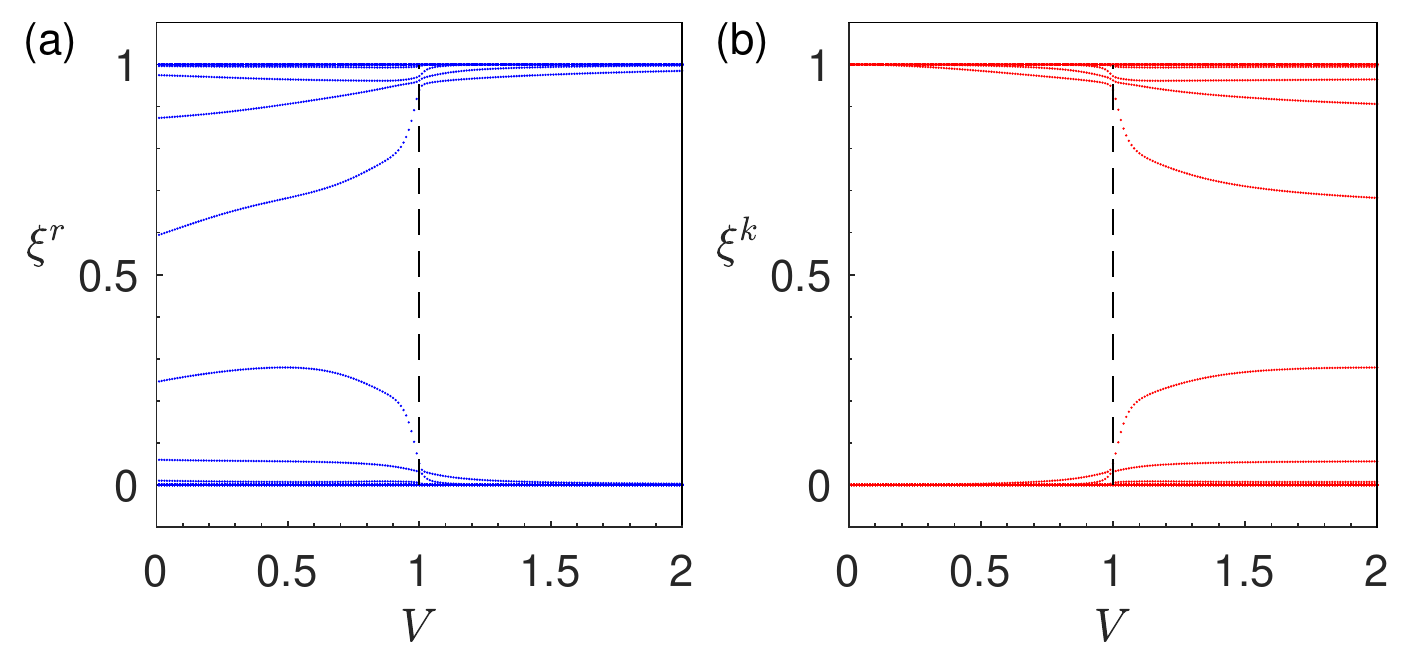}
\caption{(a) and (b) respectively show the real-space and momentum-space ES as a function of the potential strength with half-occupation, half-partition and $t=1$.}\label{AAESV}
\end{figure}

\begin{figure}[htbp]
\centering
\includegraphics[width=8.5cm]{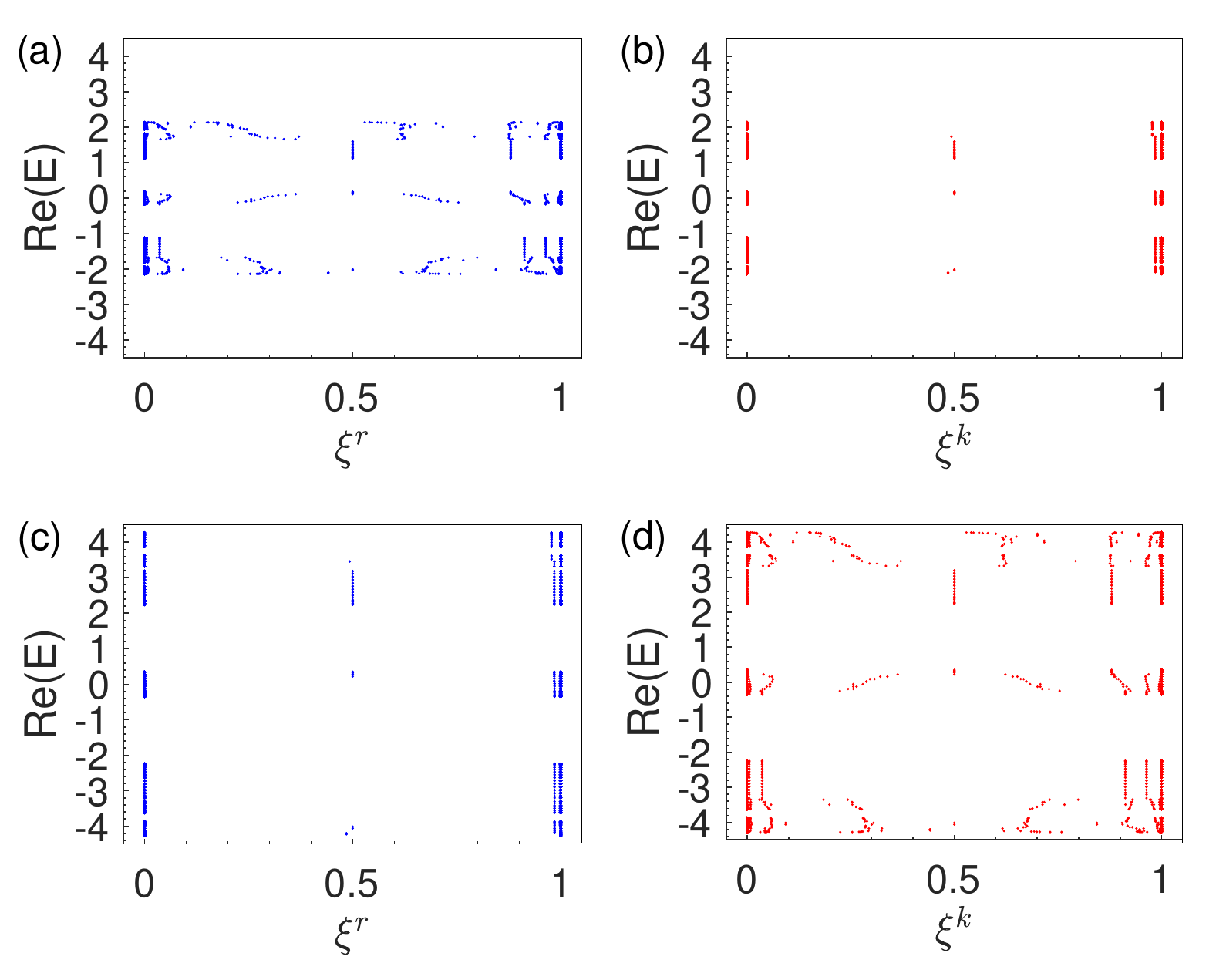}
\caption{(a) and (b) are real-space and momentum-space ES along real part of fermi energy when $t=1.0,V=0.5$ under PBC, respectively. (c) and (d) are the real-space and momentum-space ES along real part of Fermi energy when $t=1.0,V=2$, respectively. }\label{AAESE}
\end{figure}
When $J_R=J_L,V_R=V_L$, the model is the Hermitian AAH model rewritten as
\begin{subequations}
\begin{align}
E^{I}_{V} u_{n}&=t\left(u_{n+1}+u_{n-1}\right)+2V \cos (2 \pi \alpha n) u_{n} \ ,\\
\!\!E^{II}_{V} u_{m}&=2 t \cos (2\pi \alpha m) u_{m}+\!\!V\left(u_{m+1}+u_{m-1}\right) \ .\label{kE}
\end{align}
\end{subequations}
When we interpret the $m$ as lattice sites, and exchange the $t$ and $V$ in Eq.(\ref{kE}), the two model~(\ref{kE}) are also mutually duality and have a self-duality point at $t=V$.

Using similar discussion in main text, without loss of generality, we set $t=1$ and rescale Eq.(\ref{kE}), then Eq.(\ref{kE}) become
\begin{subequations}
\begin{align}
E^{I} u_{n}&=\left(u_{n+1}+u_{n-1}\right)+ 2 V \cos (2 \pi \alpha n) u_{n} \ ,\\
\frac{1}{V}E^{II} u_{m}&=\left(u_{m+1}+u_{m-1}\right)+2\frac{1}{V} \cos (2\pi \alpha m) u_{m} \,. \label{AAH3}
\end{align}
\end{subequations}
From Eq.~(\ref{AAH3}), the ES of Hermitian AAH model also have an identity between two ES eigenvalue sets:
\begin{align}
\left\{\xi^k_i(V)\right\}=\left\{\xi_i^r( {V}^{-1})\right\}\label{eqn_duality_herm}\,.
\end{align}
Finally, we choose two parameters to display the identity of Hermitian AAH model by using ES, as shown in Fig.(\ref{AAESE}) and Fig.(\ref{AAESV}).

\subsection{Details about numerical calculation}
For an incommensurate ratio $\alpha$, it has a continued fraction expansion
\begin{equation}
\alpha=n_{0}+\frac{1}{n_{1}+\frac{1}{n_{2}+\frac{1}{n_{3}+\cdots}}}=\left[n_{0}, n_{1}, n_{2}, \ldots\right]
\end{equation}
where $n_k$ are integers. Then we can use rational approximant $\alpha\approx \frac{M_k}{N_k}=\left[n_{0}, n_{1}, n_{2}, \ldots, n_k\right]$, and $N_k$ is the lattice site to satisfy periodic boundary condition. Different $n_k$ are corresponding to different $N_k$ and lead to different numbers of subbands, then the energy spectrum has a hierarchical structure. However, since the approximation to $\alpha$ changes very little, the spectrum is still dominated by the $n_1$ first-order bands, each band split into $n_2$ narrower subbands. Different $\alpha$ have the same transition point in the same model, and the scaling at transition point is sightly related to different $\alpha$\cite{szaboNonpowerlaw2018,cookmeyerCritical2020}.

In our models, we choose $\alpha=\sqrt{2}$, and it can be approximated by different fractions, such as $\{1, \frac{3}{2}, \frac{7}{5}, \frac{17}{12}, \frac{41}{29}, \frac{99}{70}, \frac{239}{169}, \frac{577}{408}, \frac{1393}{985},...\}.$ For convenient to calculate the entanglement measure, we choose $\frac{M_k}{N_k}=\frac{239}{169}$ and the lattice site is $L=169$.

In AAH model, the Fourier transformation is different from standard Fourier transformation, because AAH model has a quasiperiodicity $\frac{1}{\alpha}$, which is not a integer number and its momentum should also have the same quasiperiodicity, donated by $k=2\pi \alpha m,m\in[1,L]$. Therefore, in quasicrystal system, if we use the standard Fourier transformation to study the properties of quasicrystal model in momentum space, it would cause the misleading result.

\end{document}